\documentstyle{mn}
\begin{document}

\input{psfig.sty}
\newcommand{\pmin}{P_{\rm min}}
\newcommand{\pturn}{P_{\rm turn}}
\newcommand{\jgr}{J_{\rm GR}}
\def\lsun{{L_{\odot}}}
\def\msun{{\rm M_{\odot}}}
\def\msol{{\rm M_{\odot}}}
\def\msolyr{M_\odot {\rm yr^{-1}}}
\def\mdot{\dot M}
\def\te{T_{\rm eff}}
\def\rsun{{R_{\odot}}}
\def\be{\begin{equation}}
\def\ee{\end{equation}}
\newcommand{\porb}{P_{\rm orb}}
\newcommand{\renv}{R_{\rm env}}
\def\mmin{m_{\rm min}}
\def\xc{X_{\rm c}}
\def\mv{M_{\rm V}}
\def\te{T_{\rm eff}}
\def\lsol{L_\odot}
\def\m2i{M_{2,\rm i}}
\newcommand{\simgr}{\ga}
\newcommand{\simle}{\la}
\newcommand{\lta}{\la}
\newcommand{\gta}{\ga}
\input{epsf.sty}
\def\plotone#1{\centering \leavevmode
\epsfxsize=\columnwidth \epsfbox{#1}}
\def\plottwo#1#2{\centering \leavevmode
\epsfxsize=.45\columnwidth \epsfbox{#1} \hfil
\epsfxsize=.45\columnwidth \epsfbox{#2}}
\def\plotfiddle#1#2#3#4#5#6#7{\centering \leavevmode
\vbox to#2{\rule{0pt}{#2}}
\includegraphics{#1}}


\title{On the late spectral types of cataclysmic variable secondaries}

\author[I. Baraffe, U. Kolb]{I.~Baraffe$^{1}$ and U.~Kolb$^{2}$ \\ 
$^1$  Ecole Normale Sup\'{e}rieure de Lyon, C.R.A.L.\ (UMR 5574 CNRS),
F-69364 Lyon Cedex 07, France (ibaraffe@ens-lyon.fr)\\ 
$^2$ Department of Physics \& Astronomy, The Open University, Walton Hall,
Milton Keynes, MK7~6AA (u.c.kolb@open.ac.uk)
}



\date{Submitted to MNRAS: 15 February 2000}


\maketitle

\begin{abstract}
We investigate why the spectral type of most cataclysmic variable (CV)
secondaries is
significantly later than that of a ZAMS star with the same mean
density. 
Using improved stellar input physics, tested against observations of
low--mass stars at the bottom of the main sequence, we calculate the
secular evolution of CVs with low--mass
donors. We consider sequences with different mass transfer rates and
with a different degree of nuclear evolution of the donor prior to
mass transfer.

Systems near the upper edge of the gap ($P \sim 3 - 6$ h)
can be reproduced by models with a wide range of mass transfer rates
from $1.5 \times 10^{-9} \, \msolyr$ to  $10^{-8} \, \msolyr$. Evolutionary
sequences with a small transfer rate and donors that are substantially
evolved off the ZAMS (central hydrogen content $0.05-0.5$) reproduce
CVs with late spectral types above $P \simgr$ 6 h. Systems with the
most discrepant (late) spectral type should have the smallest donor
mass at any given $P$. 

Consistency with the period gap suggests that the mass transfer rate
increases with decreasing donor mass for evolved sequences above the
period gap. In this case, a single--parameter
family of sequences with varying $\xc$ and increasing mass transfer
rate reproduces the full range of observed spectral types.
This would imply that CVs with such evolved secondaries
dominate the CV population. 
\end{abstract}
 

\section{Introduction}

Over last two decades, numerous studies have considered the evolution of
the low--mass secondaries in cataclysmic variable (CV) systems. The
calculations 
aimed mainly at reproducing the observed distribution of orbital
periods $P$ of CVs, in particular the minimum period at 80 min and
the dearth of systems in the 2-3 h period range (see e.g.\ King 1988
for a review). Based on the most popular explanation for this period
gap, the disrupted magnetic braking model (cf.\ Rappaport, Verbunt and
Joss 1983; Spruit and Ritter 1983), 
evolutionary sequences constructed with full stellar models or
simplified bipolytrope models (cf. Hameury
et al. 1988; Hameury 1991; Kolb and Ritter 1992; and references
therein) reproduce the broad features of the observed $P$ distribution.
Despite the significant uncertainties in the underlying stellar input
physics (e.g.\ opacities, equation of state, etc...), the lack of a reliable
description of angular momentum losses $\dot J$ 
that drive the mass transfer provided sufficient
freedom that it was always possible to find a set of secondary star models which
reasonably fit the period gap.  

Input physics like atmospheric opacities and the outer
boundary condition 
are essential for a comparison with observable
quantities such as colours or spectral types. 

The spectral type and the orbital period ($P \propto (R^3/M)^{1/2}$; $R$ and $M$
 are the secondary radius and mass) provide  
a unique set of constraints that test the structure both of the secondary's
interior and its outermost layers.
As we shall show below, these can be used to find
constraints on the actual orbital braking strength $\dot J$ as well. 

Previously, an analysis of this kind was hampered by the lack of
knowledge, both theoretical and observational, of stars in the
low--mass range populated by CV secondaries. 
Here we make use of the marked and independent progress in the study of
low--mass stars over the past decade. 
A wealth of ground-based and spaced-based observations has
significantly improved our knowledge of the photometric and
spectroscopic properties of stars at the bottom of the 
main sequence. 
In parallel, recent theoretical work has demonstrated the necessity
to use accurate internal physics and outer boundary conditions based on
non--grey atmosphere models to correctly describe the structure and
evolution of low--mass objects (Burrows et al. 1993;  Baraffe et
al. 1995, 1997, 1998; Chabrier and Baraffe 1997, 2000,  and references
therein). These interior models, 
combined with recent, much improved models of cool atmospheres
and synthetic spectra (see the review of Allard et al.\ 1997, and
references therein) allow one 
to calculate self--consistent magnitudes and synthetic colours
of low--mass stars. These can be
compared directly to observed quantities, avoiding the 
use of uncertain empirical effective temperature $\te$ and 
bolometric correction scales.  
The agreement between models and observations for
(i) eclipsing binary systems (Chabrier and Baraffe 1995), (ii)
mass--magnitude and mass--spectral type relations (Baraffe and
Chabrier, 1996; Chabrier et al. 1996), (iii) colour--magnitude diagrams
(Baraffe et al. 1997, 1998) and (iv) synthetic spectra (Leggett et
al. 1996; Allard et al. 1997), demonstrate how reliable the theory
of low--mass stars already is.

Recently, Beuermann et al.\ (1998) used our preliminary 
calculations based on these next--generation low--mass star
models and compared calculated and observed spectral types
(SpT) of CV secondaries.
Beuermann et al.\ (1998) found that CVs with donors that are nuclearly
evolved, with a substantial amount of central hydrogen depletion, can
account for the late spectral types in long--period systems ($P/{\rm h}
\simgr 6$), but not in those close 
to the upper edge of the period gap ($3 \simle P/{\rm h} \simle 6$).
Higher mass transfer rates as usually assumed
can account for the latter. 

The purpose of the present paper is to give a detailed account of the
full set of our CV evolutionary calculations, a preliminary subset of 
which has been used by Beuermann et al.\ 1998. We discuss
implications of the need for both high mass transfer rates and
a significant fraction of systems with nuclearly evolved donors.
We focus on CVs where the spectral type can be
determined fairly accurately, i.e.\ on systems with orbital period
above 3 h. In a previous paper (Kolb and Baraffe 1999), we 
considered aspects specific to the evolution of systems below the
period gap ($P \le 2$ h).   
  
In \S 2 we summarize the input physics and present 
evolutionary sequences with initial ZAMS donors.
Nuclearly evolved sequences are presented in \S 3. A discussion and 
conclusions follow in \S 4.

\section {Evolutionary models with initial ZAMS donor}

A brief description of the input physics for our stellar code
can be found in Kolb and Baraffe (1999) and Baraffe et al.\ (1998).
More details are given by Chabrier and Baraffe (1997)
for the interior physics and by Hauschildt et al.\ (1999) for the
atmosphere models. 
Here we just recall the main improvements compared to earlier
models applied to CV donors: (i) the new models are
based on the equation of state (EOS) of Saumon et al.\ (1995) which is
specially devoted to low--mass stars and brown dwarfs, and to the
description of strong non--ideal effects characterizing the interior
stellar plasma; (ii) the outer boundary condition is based on
non-grey atmosphere models (the use of a grey boundary condition overestimates the
effective temperature $\te$ and luminosity $L$ 
for a given mass $M$); 
(iii) mass--colour and mass--magnitude relations are derived
self--consistently from the synthetic spectra of the same atmosphere
models which provide the boundary condition.

We use the empirical SpT -- $(I-K)$ relation established by Beuermann et
al.\ (1998) to determine the spectral type SpT for a given model ($M$,
$\te$, $L$) from the calculated colour $I-K$.
The accuracy and relevance of this conversion is discussed in Beuermann
et al.\ (1998). 
 
Because of the uncertainty of current descriptions for orbital angular 
momentum losses by magnetic braking
and the substantial computer time (cf. Kolb and Ritter, 1990;
Hameury 1991) required for evolutionary sequences if mass transfer is
treated explicitly and self--consistently,
we restrict our analysis to sequences calculated with constant
mass transfer rate.  
We have checked that 
sequences with mass transfer driven by angular momentum losses 
$\dot J$ with either
constant $\dot J$, constant $\dot J/J$, or $\dot J$ according to Verbunt \&
Zwaan (1981), have the same main properties as the sequences with 
constant $\dot M$ on which we base our conclusions.

\begin{figure*}
\psfig{file=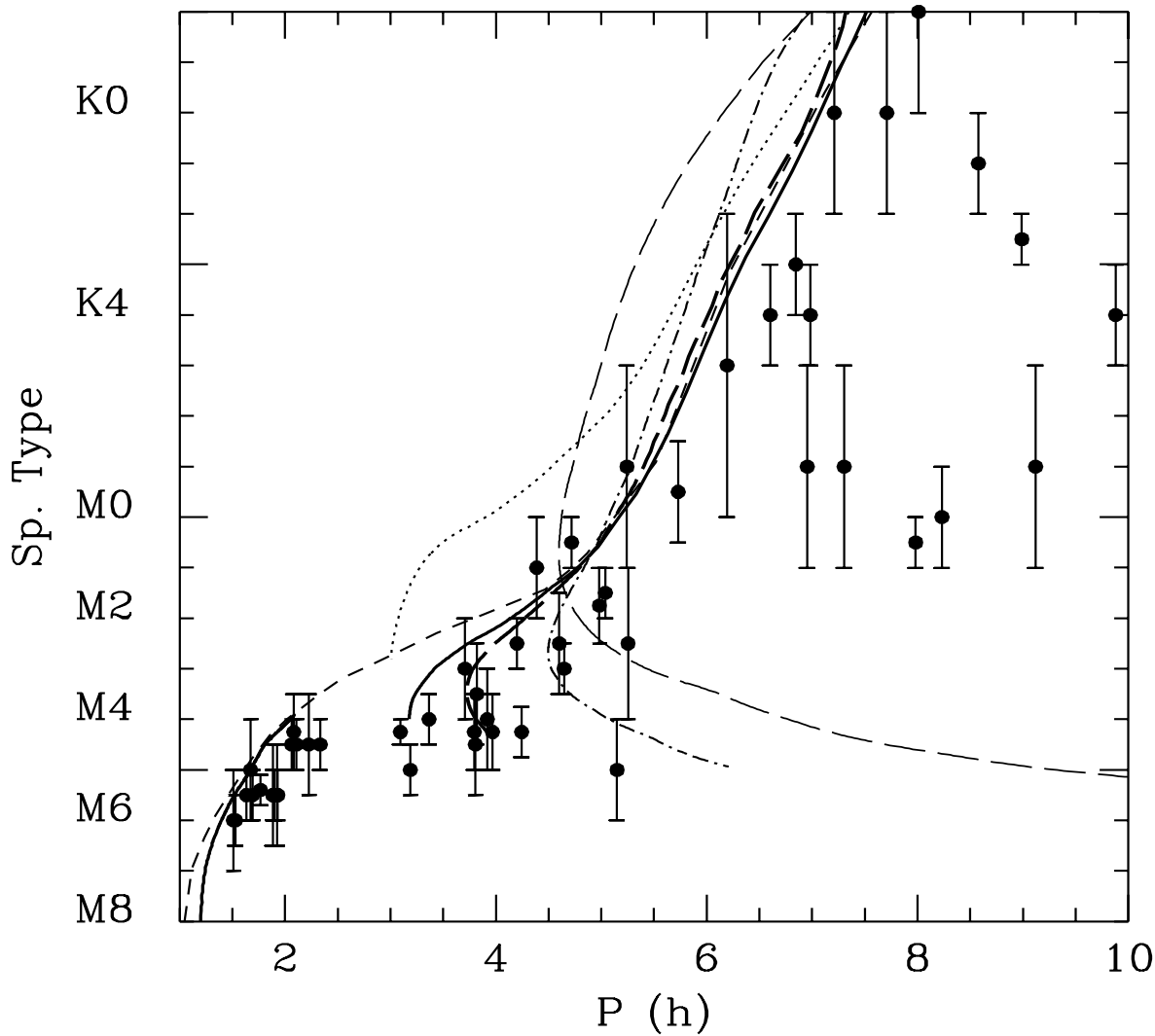,height=160mm,width=170mm}
\caption{Spectral type of CV secondaries versus orbital period, for 
evolutionary tracks with unevolved donor stars and different 
mass transfer rate. 
The observed data are taken from Beuermann et al.\ (1998), with slight updates. 
The short--dashed curve corresponds to
solar metallicity stellar models on the ZAMS. The thick solid curve is
the standard sequence with constant $\dot M = 1.5 \times 10^{-9} \msolyr$
above the period gap and $\dot M = 5 \times 10^{-11} \msolyr$ below the gap. The
thick long--dashed curve is with constant $\mdot$ = $3 \times 10^{-9}$;  
the dash--dotted curve is with  $\mdot= 10^{-8} \msolyr$;
the thin long--dashed curve
is with $\mdot= 10^{-7} \msolyr$. For all sequences, the initial secondary
mass is $M_2$ = 1 $\msun$.
Sequences above the period gap
terminate when the donor becomes fully convective. 
The dotted curve corresponds to a sequence with $\mdot
= 1.5 \times 10^{-9} \msolyr$, but calculated with the Eddington
approximation as outer boundary condition for the stellar models.
}
\label{fig1}
\end{figure*}

In this section we consider sequences where mass transfer starts 
with donors of mass $M_2 = 1 \msol$ that are initially on the ZAMS, 
with solar composition (X = 0.70, Z=0.02). 
We define a ``standard sequence'' 
which reproduces the width and location of the period gap (2.1 - 3.2 h) in 
the framework of the disrupted magnetic braking model.
This sequence is calculated with $\mdot = 1.5 \times 10^{-9} \msolyr$ until
the donor becomes fully convective at $M_2$ = 0.21 $\msol$ and $P$ =
3.2 h. At this point mass transfer ceases and the secondary shrinks back to
its thermal equilibrium configuration in  $2-3 \times 10^8$ years.
Mass transfer resumes when the Roche lobe radius catches up with the
donor radius at a period $P$ = 2.1 h. The transfer rate in this second 
phase is $5 \times 10^{-11} \msolyr$, a value typical for systems driven by
gravitational wave emission (cf.\ Kolb and Baraffe 1999). 
This standard sequence is shown in Fig. \ref{fig1} (thick solid line),
together with sequences calculated with higher transfer rates ($3
\times 10^{-9} \msolyr$, $10^{-8} \msolyr$ and $ 10^{-7} \msolyr$). 

\begin{figure*}
\psfig{file=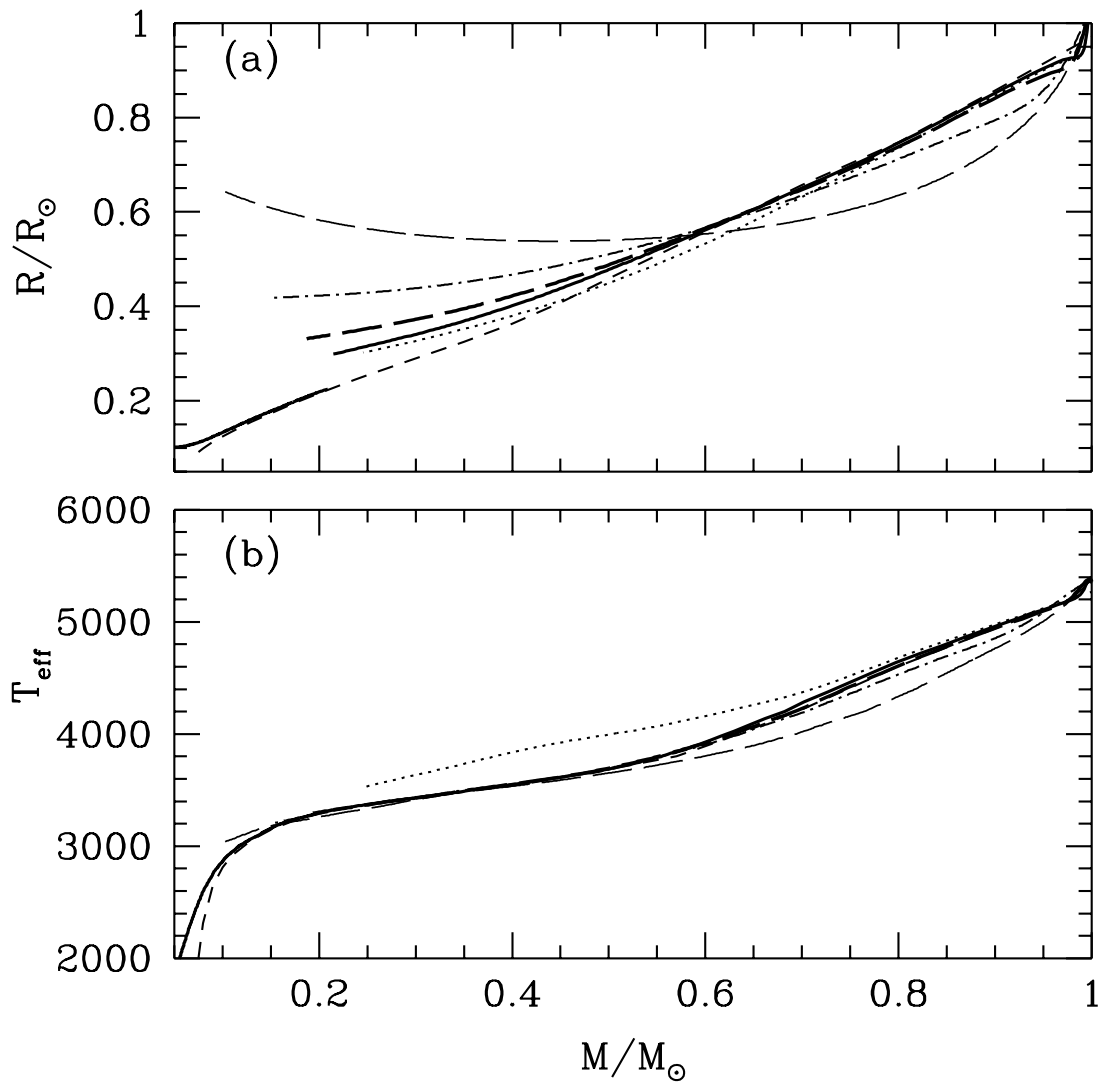,height=160mm,width=160mm}
\caption{{\bf (a)} Mass -- radius relation for ZAMS stars and
for unevolved CV donors in sequences with constant mass transfer rate. Linestyle 
as in Fig.~1. {\bf (b)} Mass -- $\te$ relation for the same
models as in (a). 
}
\label{fig2}
\end{figure*}

Mass--losing stars deviate from thermal equilibrium. The deviation is
large if the secondary's thermal timescale $t_{\rm KH} \sim
GM^2/RL$ is long compared to the mass transfer timescale 
$t_M = M/(-\dot M)$. As shown in Fig. \ref{fig2}a, mass transfer
causes a contraction of the donor radius (e.g.\ Whyte and
Eggleton 1980; Stehle et al.\ 1996) relative to the corresponding ZAMS
radius for predominantly radiative objects $M \simgr 0.6 \msol$, where the
radiative core exceeds 80\% of the total mass (cf.\ Chabrier and Baraffe 1997,
their Fig.~9). Predominantly convective stars ($M \simle 0.6 \msol$)
expand relative to the ZAMS. Although a large mass transfer rate
causes a significant departure of R from its equilibrium
value (see Fig. \ref{fig2}a),  Fig. \ref{fig2}b shows that $\te$ is
rather insensitive to $\mdot$ (see also 
Singer et al.\ 1993; Kolb et al.\ 2000). 

The outer boundary condition of the stellar model is more important
than $\dot M$ for the determination of $\te$, as indicated by the
dotted curve in Fig.~\ref{fig1} and Fig.~\ref{fig2}b.
This curve corresponds to a sequence with $\mdot = 1.5 \times 10^{-9}
\msolyr$, but calculated with the Eddington
approximation, which implicitly assumes greyness and the diffusion
approximation for radiative transfer in the atmosphere. Chabrier and
Baraffe (1997)  
have shown that such an approximation overestimates $\te$ for
a given mass when molecules form in the atmosphere 
at $\te \simle 4000 - 4500$~K. 
Note that the Eddington approximation sequence reproduces
the period gap as well as the corresponding non-grey
sequence, with a width and location of 2.2 - 3 h. However, the Eddington
sequence consistently gives earlier spectral types than observed, and 
dramatically 
fails to reproduce the location of objects with $P \le
5$ h in Fig. \ref{fig1}. 
This example highlights the significant improvements of 
our new generation of low--mass star models.

Note that the period $\pturn$ where the period derivative changes from 
negative to positive due to departure from thermal equilibrium 
increases with mass transfer rate.
Remarkably, the standard sequence reaches period bounce at
$P$ = 3.2 h, the same period where the secondary star becomes
fully convective. Sequences with higher mass transfer rate bounce  
before the donor becomes fully convective. 
In other words, for mass transfer rates $\mdot \ge 
1.5 \times 10^{-9} \msolyr$, the minimum period reached by CVs
just corresponds to or exceeds the observed upper edge of the period
gap at $\sim 3$ h.  

As mentioned in Beuermann et al.\ (1998), a large spread of the secular
mean mass transfer rates from $1.5 \times 10^{-9} \msolyr$ to $\sim
10^{-8} \msolyr$ could account for the observed range of late
spectral type objects with 
$P \simle 5$ h. It is well known that such a spread of $\dot M$
in systems above the period gap cannot be excluded by observations
(cf.\ e.g.\ Warner 1995, his Fig.~9.8; Sproats et al.\ 1996).

Although the sharp observed gap boundaries seem to
imply uniform values of the secular mean $\mdot$ near the
upper edge of the period gap (e.g.\ Ritter 1996), a spread 
of $\mdot$ cannot be dismissed unambiguously. The sharp 
upper edge is preserved if disrupted magnetic braking holds
{\em and} $\mdot$ adopts only values $\geq 1.5 \times 10^{-9} \msolyr$. 
With increasing $\mdot$ the critical mass 
where the secondary becomes fully convective becomes smaller 
(cf.\ Fig.~\ref{fig2}a).  
Thus, if the system detaches at this point, by analogy with the
disrupted magnetic braking model, mass 
transfer resumes after the secondary has reached thermal
equilibrium at a period shorter than 2 h. As an example, for 
the sequence with $\mdot =  10^{-8} \msolyr$ the donor becomes fully
convective at $M_2 \sim 0.15 \msol$, and the system
reappears as CV at $P \sim 1.7$~h.  
Test calculations with a population synthesis code (e.g.\ Kolb 1993) 
show that such a spread of systems reappearing below the gap 
still gives a reasonably sharp lower edge of the period gap,
with only a mild over--accumulation of systems in a period interval 
$1.7-2.1$~h. This is perfectly consistent with the observed
distribution which does show a slight, although statistically probably
not significant, accumulation of systems there. 

Despite this consistency, accepting the $\mdot$ spread is not an attractive 
solution to the late SpT problem. It is hard to see why a presumably 
global 
angular momentum loss mechanism should drive such different transfer
rates in otherwise similar systems. 

Finally we note that there are no CV secondaries above the ZAMS line in the 
$P-$SpT diagram 
for periods longer than $\sim 6$~h. This suggests that in these systems
the secular mean $\dot M$ must be smaller than $10^{-8} \msolyr$.   
But whatever the value of $\mdot$, sequences with unevolved donors
never drop below the ZAMS line for $P \ga 6$~h.

\begin{table}
\caption{Donor mass $M$ and mass $M_{\rm core}$ of its radiative core, 
at $P$ = 3 h for evolved sequences with initial donor mass $M_2 = 1$ or 1.2 $\msol$,  
constant mass transfer $\mdot = 1.5 \times 10^{-9} \msolyr$ and 
different central H abundance $\xc$ at the time mass transfer
starts.  $t_0$ is the age of the donor when mass transfer turns on. 
}
\begin{tabular}{lcccc}
$M_2/\msol$ & $\xc$ & $t_0$ (Gyr) & $M/\msol$ & $M_{\rm core}/\msol$ \\
 &&&& \\
1 & 0.55 & 2.37 & 0.191 & 0.064 \\ 
  & 0.16 & 7.2 & 0.209 & 0.125 \\
  & 0.05& 8.16 & 0.199 & 0.136 \\
 &&&& \\
1.2 & 0.16 & 3.75 & 0.183 & 0.120 \\
    & 0.05 & 4.18 & 0.143 & 0.127 \\
  \end{tabular}
  \end{table}

\section{Evolved sequences}

We now consider sequences with initial donors which have
evolved off the ZAMS prior to mass transfer. When mass transfer 
turns on,
the star is old enough (cf.\ Table 1) to have burned a substantial
amount of hydrogen in the centre. Only stars with mass $M > 0.35 \msol$,
which develop a radiative core (cf.\ Chabrier and Baraffe 1997),
start to deplete central H within a Hubble time. But only
if $M > 0.8 \msol$ is more than $50\%$ of the initial H depleted in less
than 10 Gyr. Therefore, this donor type is restricted to systems that 
form with a fairly massive secondary $\ga 1 \msun$.

\begin{figure*}
\psfig{file=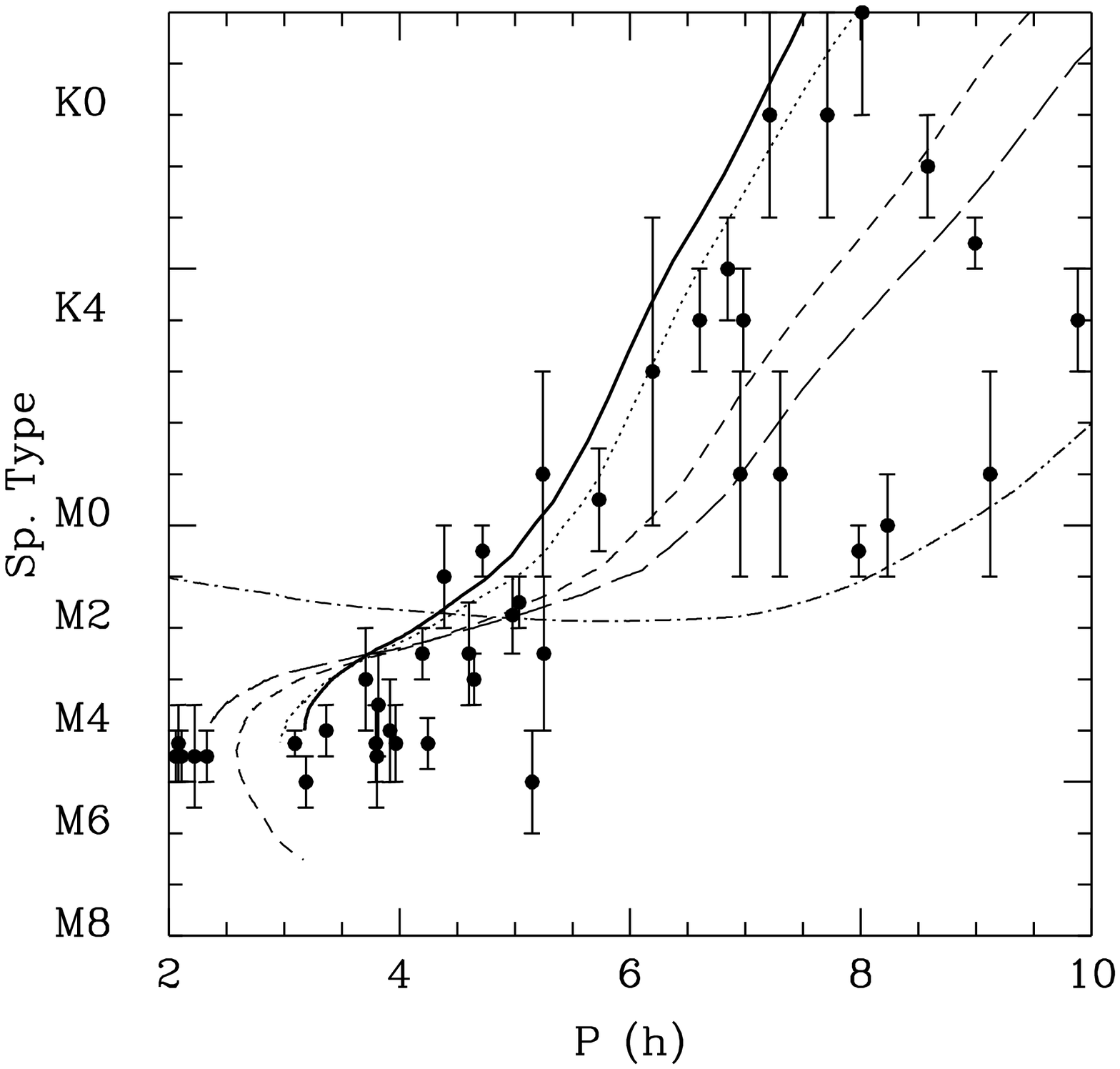,height=160mm,width=160mm}
\caption{As Fig.~1, but for sequences with different initial
central H abundance $X_{\rm c}$ at the time mass transfer starts.
The mass transfer rate is $\mdot = 1.5 \times
10^{-9} \msolyr$, the initial secondary mass $M_2 = 1$ and 1.2  $\msol$.
Thick solid line: standard
sequence with $M_2 = 1 \msol$ and $X_{\rm c} = 0.7$; dotted curve: $M_2 = 1 \msol$
and $X_{\rm c} = 0.55$; dashed
curve: $M_2 = 1 \msol$ and $X_{\rm c} = 0.16$; long--dashed curve: $M_2 = 1 \msol$ and
$X_{\rm c} = 0.05$;
dash--dotted curve: $M_2 = 1.2 \msol$ and $X_{\rm c} = 0.05$. 
The data are the same as in Fig.~\ref{fig1}.}
\label{fig3}
\end{figure*}

Figure~\ref{fig3} depicts sequences with such nuclearly evolved
donors --- hereafter ``evolved sequences'' --- in the
$P$--SpT diagram, for different central H mass fractions $X_{\rm c}$ at
the time mass transfer starts. 
The initial secondary mass is $M_2 = 1 \msol$ or $M_2 = 1.2 \msol$ and 
mass transfer rate is fixed to $1.5 \times 10^{-9} \msolyr$. 
Table 2 gives for each of these sequences
the secondary mass $M$, spectral type, $\te$ and
radius $R$ as a function of $P$, for $3 \le P/h \le 10$.
The evolutionary track of ZAMS donor sequences in the $P$--SpT diagram 
is insensitive to the initial donor mass, while
for evolved sequences with the same initial $X_{\rm c}$ $\te$
decreases (the spectral type becomes later) at a given $P$ with increasing 
initial donor mass.
This is illustrated in Fig.~\ref{fig3} where we plot 
sequences starting with $M_2 = 1 \msol$ (long--dashed curve) and $M_2 = 1.2 \msol$ 
(dash--dotted curve), both for $X_{\rm c} = 0.05$.

\begin{figure}
\psfig{file=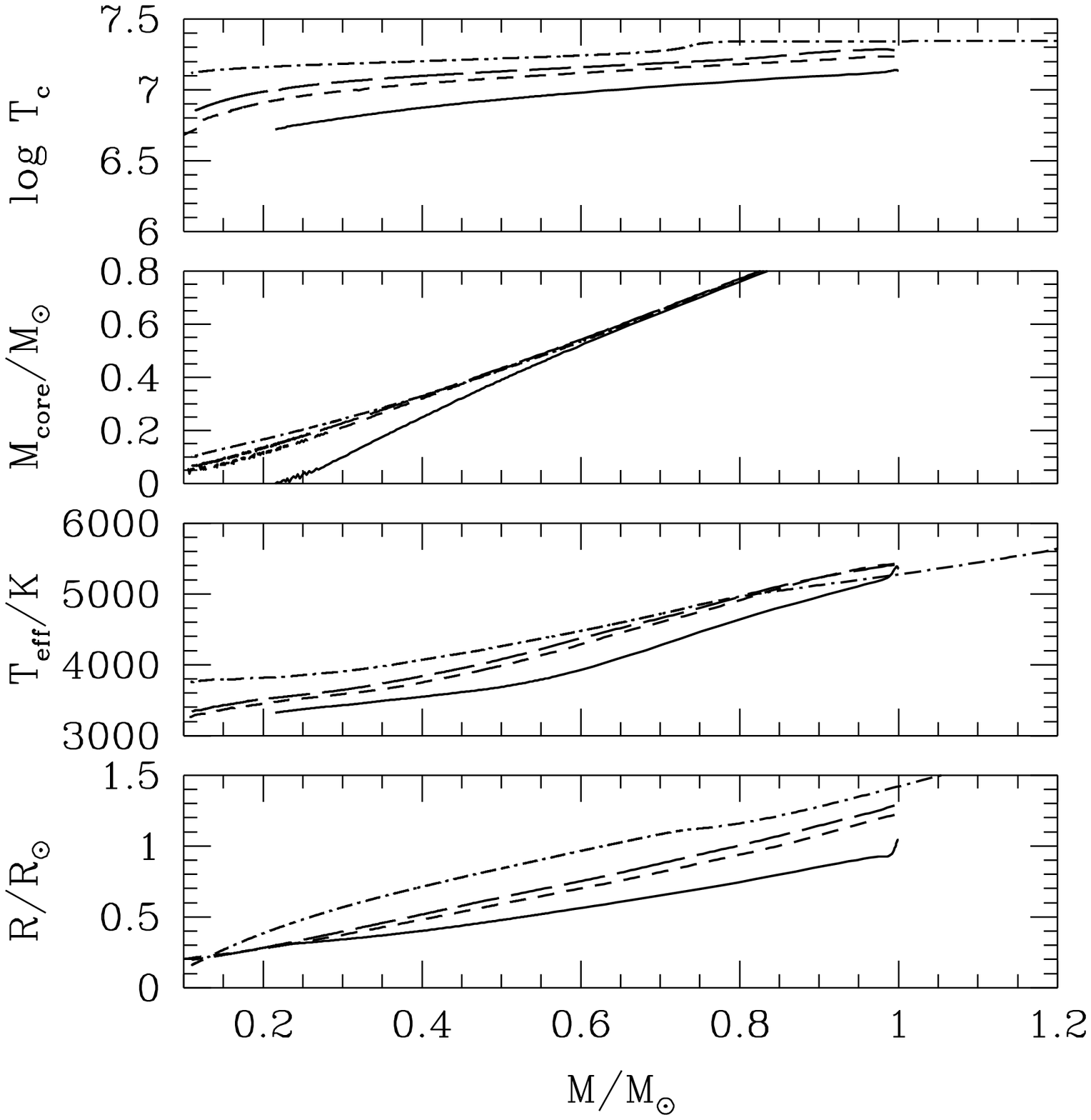,height=160mm,width=88mm}
\caption{Variation of several stellar quantities with mass for
sequences with constant mass transfer rate $\mdot = 1.5 \times 10^{-9} \msolyr$,
 initial masses $M_2 = 1$ and 1.2 $\msol$ and different central H abundances
$X_{\rm c}$ at the time mass transfer starts.
Solid curve: standard
sequence with $M_2 = 1 \msol$ and $X_{\rm c} = 0.7$;  dashed curve: $M_2 = 1 \msol$
and $X_{\rm c} = 0.16$; long--dashed curve: $M_2 = 1  \msol$ and $X_{\rm c} = 0.05$;
dash--dotted curve: $M_2 = 1.2  \msol$ and $X_{\rm c} = 0.05 $. 
$M_{\rm core}$ is the mass
of the radiative core. $T_{\rm c}$ is the central temperature (in K).
}
\label{fig4}
\end{figure}

\begin{figure}
\psfig{file=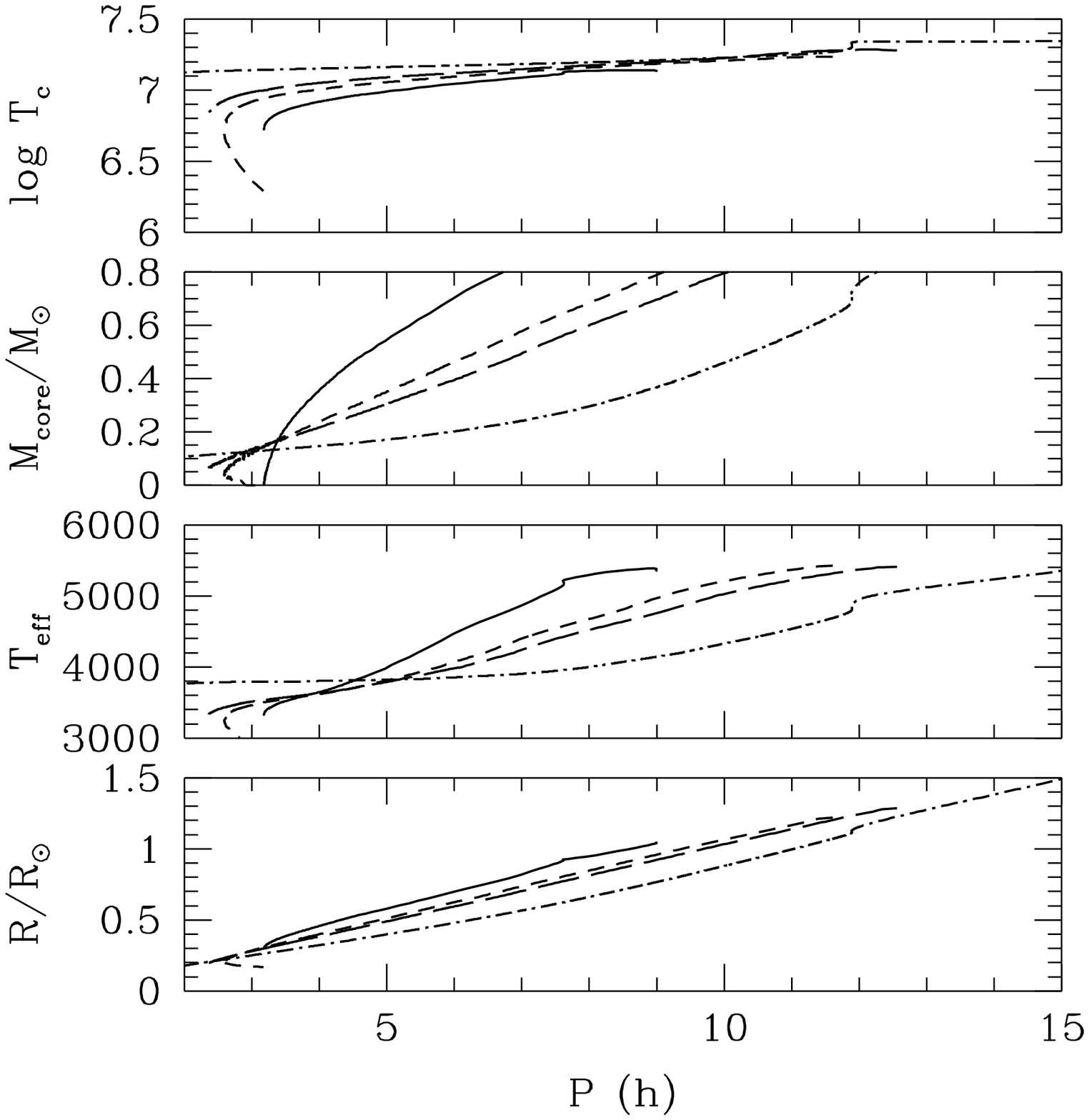,height=160mm,width=88mm}
\caption{Same as Fig.~4, but as a function of orbital period.
}
\label{fig5}
\end{figure}

As already emphasized by Beuermann et al.\ (1998), evolved
sequences result in later spectral types for a given 
$P \ga 5$~h than
the standard sequence and can explain amazingly 
well the spread of data observed above 6 h. 
The reason for this is illustrated in Figures \ref{fig4} and
\ref{fig5}, which display several diagnostic
quantities along the sequences as a function of secondary mass
(Fig.~\ref{fig4}) and orbital period (Fig.~\ref{fig5}).  

When mass transfer turns on, more evolved donors
with smaller $X_{\rm c}$ have larger $R$, $\te$ and $T_{\rm c}$ than 
less evolved donors.
This follows from the well known properties of the H burning phase 
of solar--type stars with a central radiative core: as hydrogen is
depleted in the centre, the increase of the central molecular weight
$\mu$ yields an increase of the central temperature and thus $L$, 
causing an expansion of the star. Hence the main--sequence evolution of 
a star with constant mass proceeds towards larger $L$ and $\te$. 
The same effect exists for mass--losing main--sequence stars. 
Consequently, {\it for a given mass}, $R$ is larger for more
evolved sequences, and $P$ is longer, while {\it for a given $P$},  
the secondary mass and $\te$ is smaller, hence the spectral type is later  
(cf.\ Fig.~\ref{fig3}). As a consequence, the donor mass at any given
$P$ is smallest for the most evolved sequence, cf.\ Table~2.

Although the evolved sequences shown in Fig~\ref{fig3}
provide an attractive explanation for the observed data above 5-6 h,
they are obviously in conflict with the standard explanation of the
period gap. Indeed, with the adopted value $\mdot = 1.5 \times 10^{-9}
\msolyr$ we find that these sequences enter the period
gap before the donor becomes fully convective or before the system
bounces due to departure from thermal equilibrium. Table~1 gives the mass
of the secondary and the mass $M_{\rm core}$ of its radiative core
at $P$ = 3 h for the different sequences. 
Period bounce occurs for
the ($M_2 = 1 \msun$, $\xc = 0.16$) sequence at
$P$ = 2.60 h, corresponding to $M = 0.1 \msol$ and
$M_{\rm core} = 0.04 \msol$. The donor becomes fully convective
only at $M \sim 0.05 \msol$ ($P$ = 2.9 h). For more evolved sequences,
period bounce would occur at even smaller periods. 

This difference in the size of the radiative core
is caused by the small central H abundance.  
The smaller $\xc$ and the higher central temperature of evolved donors
compared to unevolved donors (cf.\ Figs.~\ref{fig4} and \ref{fig5}) 
imply lower radiative opacities in the central
region, and consequently favour radiative transport.
As a result, the inward progression of
the bottom of the donor's convective envelope as mass decreases
proceeds at a slower rate.
Hence in evolved
sequences the donor has a larger radiative core $M_{\rm core}$ than 
a donor with the same mass in an unevolved sequence (see
Fig.~\ref{fig4}). In terms of $P$, Fig.~\ref{fig5} shows that
$M_{\rm core}$ is larger for the standard sequence
at any given  $P \simgr 3.5 $ h. The situation then reverses at
the upper edge of the gap, which corresponds to $M_{\rm core} = 0$
for the standard sequence. 
The effect of a lower central H abundance described above is similar to that 
reported by Laughlin et al. (1997) who followed the central H
burning phase of very low--mass stars with $M \le 0.25 \msol$ (with
constant mass). They   note that stars which are fully convective on
the ZAMS develop a radiative core towards the end of the main sequence
because of the depletion of H and subsequent lowering of the central
radiative opacities.   

The $P$--SpT diagram 
suggests that at long orbital periods a
fairly large fraction of systems has a significantly evolved donor. 
If these carried on evolving with a transfer rate 
$\simeq 1.5 \times 10^{-9}
\msolyr$ they would clearly overpopulate the period gap regime. 
This is made worse by the fact that some of these sequences reach their 
minimum period in
the period gap (Fig.~\ref{fig3}): at period bounce $\dot P$ = 0,
and the detection probability $d \propto 1/\dot P$ increases sharply.
We find a similar behaviour if 
mass transfer is treated self--consistently, with an angular
momentum loss rate $\dot J =$~const.\ or $\dot J \propto J$,
calibrated such that the corresponding unevolved sequence reproduces
the observed period gap (giving $\dot M = 1.5 \times 10^{-9}
\msolyr$ at $P \sim 3$~h).

An obvious way to prevent evolved sequences from evolving
into the period gap region is to increase the mass transfer rate. 
This increases the deviation from thermal equilibrium and leads to 
period bounce above the upper edge of the observed gap. 
Test calculations with constant mass transfer rate show that period
bounce occurs at $P > 3$ h with $\mdot \ge 3 \times 10^{-9}$ and $ 5 \times
10^{-9}$ 
 $\msolyr$ for sequences with respectively
$\xc = 0.16$ and  0.05,
for initial donor mass
$1\msun$. 
The same behavior is found for donors with higher initial mass.
In other words, for sequences with smaller $\xc$ a higher 
$\mdot$ {\it near the upper edge} of the period gap is required than for 
less evolved donors, consistent with the fact that the thermal time is shorter 
if $\xc$ is small.  
These evolved sequences with high transfer rates ($\mdot > 5
\times  10^{-9} \, \msolyr$) also match the observed spectral types in the
region $P$ = 3-6 h, just as high $\mdot$ sequences with ZAMS donors do 
(see \S2 and Fig.~\ref{fig1}).

Note however, that for long periods $P$ $\simgr 5$ h a higher
$\mdot$ has the 
same effect as for the unevolved sequences discussed in \S 2
(Fig.~\ref{fig1}). The  
effective temperature is larger and therefore the spectral type
earlier than for a sequence with smaller $\mdot$. Thus in order to
account for the full range of rather cool spectral types at long
periods, $\mdot$ cannot always be high along all evolved
sequences.  

To summarize: evolved sequences require high $\mdot$
($ \simgr  5 \times 10^{-9}$) near the upper gap region to avoid
crossing the gap, but a low mass transfer rate 
($\mdot \simle  5 \times 10^{-9}$) for $P \simgr 6$ h to explain the late
spectral type CVs in this region. 
This suggests that the mean mass transfer rate increases during the
secular evolution of nuclear evolved donors.

\begin{table}
\caption{Characteristic quantities as a function of $P$  (for $3 \le P/h
\le 10$) for
sequences with initial donor mass $M_2 = 1$ or 1.2 $\msol$,  
constant mass transfer rate $\mdot = 1.5 \times 10^{-9} \msolyr$ and 
different central H abundance $\xc$ at the time mass transfer
starts. 
}
\begin{tabular}{lllllll}
$M_2$ & $\xc$ & $P$ & $M$ & $SpT$ & $\te$ & $R$ \\
$\msol$ & & h & $\msol$ & & K & $R_\odot$ \\
 &&&&&& \\
  1.00&  0.700&  9.0&  1.000&  G5-G6&  5360.&  1.0441\\
  & & 8.0&  0.992&  G6&  5289.&  0.9454\\
  & &  7.0&  0.870&  K0-K1&  4870.&  0.8205\\
  & &  6.0&  0.751&  K4-K5&  4467.&  0.6972\\
  & &  5.0&  0.616&  M0-M1&  3979.&  0.5758\\
  & &  4.0&  0.475&  M2-M3&  3648.&  0.4574\\
  & &  3.5&  0.377&  M3&  3521.&  0.3856\\
 &&&&&& \\
  1.00&  0.550&  9.0&  0.976&  G6&  5276.&  0.9933\\
  & & 8.0&  0.912&  G8&  5105.&  0.9111\\
  & &  7.0&  0.795&  K1-K2&  4713.&  0.7784\\
  & &  6.0&  0.684&  K6&  4293.&  0.6666\\
  & &  5.0&  0.571&  M1&  3905.&  0.5626\\
  & &  4.0&  0.437&  M2-M3&  3632.&  0.4433\\
  & &  3.0&  0.191&  M4&  3324.&  0.2764\\
 &&&&&& \\
  1.00&  0.160& 10.0&  0.891&  G7&  5201.&  1.0617\\
  & & 9.0&  0.815&  G9-K0&  4961.&  0.9571\\
  & &  8.0&  0.724&  K2-K3&  4672.&  0.8456\\
  & &  7.0&  0.629&  K5-K6&  4385.&  0.7324\\
  & &  6.0&  0.530&  M0-M1&  4071.&  0.6239\\
  & &  5.0&  0.427&  M1-M2&  3809.&  0.5104\\
  & &  4.0&  0.325&  M2-M3&  3619.&  0.3998\\
  & &  3.0&  0.209&  M3-M4&  3464.&  0.2848\\
 &&&&&& \\
  1.00&  0.050& 10.0&  0.820&  G8-G9&  5025.&  1.0309\\
  & &  9.0&  0.735&  K1-K2&  4756.&  0.9215\\
  & &  8.0&  0.649&  K4&  4516.&  0.8113\\
  & &  7.0&  0.554&  K7&  4232.&  0.7000\\
  & &  6.0&  0.463&  M1&  3981.&  0.5941\\
  & &  5.0&  0.376&  M1-M2&  3789.&  0.4884\\
  & &  4.0&  0.288&  M2-M3&  3627.&  0.3820\\
  & &  3.0&  0.199&  M3&  3514.&  0.2800\\
 &&&&&& \\
  1.20&  0.050& 10.0&  0.531&  K6&  4329.&  0.8796\\
  & &  9.0&  0.441&  K7-M0&  4145.&  0.7668\\
  & &  8.0&  0.362&  M1&  3999.&  0.6613\\
  & &  7.0&  0.299&  M2&  3905.&  0.5655\\
  & &  6.0&  0.248&  M2&  3853.&  0.4785\\
  & &  5.0&  0.206&  M2&  3821.&  0.3969\\
  & &  4.0&  0.172&  M1-M2&  3805.&  0.3221\\
  & &  3.0&  0.143&  M1-M2&  3791.&  0.2509\\
 &&&&&& \\
  \end{tabular}
  \end{table}

\section{Discussion and conclusions}

We performed calculations of the long--term evolution of CVs with
up--to--date stellar models that fit observed properties of single
low--mass stars exceptionally well. Our focus was on the 
orbital period-- spectral type ($P$--SpT) diagram, and the observed
location of CV secondaries which populate a band with spectral types 
significantly later than ZAMS stars at the same orbital period.

Our calculations tested if the observed spectral types can be reproduced 
by varying two main
parameters, the mass transfer rate $\dot M$ (assumed constant), and
the degree of nuclear evolution of the secondary before mass transfer
starts (measured in terms of the initial central H fraction $\xc$).
Only CVs forming with massive ($\ga 1-1.2\msun$) donors, i.e.\ at
long periods, can have donors where  H  is significantly depleted in the
centre.  

We summarize our results as follows: 
(i) In the framework of the standard discontinuous orbital braking 
model
for the period gap, the observed gap between 2-3~h is well reproduced
by sequences starting from ZAMS donors and proceeding at a mass
transfer rate $\mdot \sim 1-2 \times 10^{-9} \msolyr$ near the upper edge
of the gap. 
(ii) Higher transfer rates ($\mdot \simgr 5 \times 10^{-9} \msolyr$)
than assumed in (i) near the upper edge of the gap are required to
explain the full range of late spectral types of secondaries in CVs
with periods $P = 3-6$ h. This is true whatever the evolutionary state of
the donor at turn--on of mass transfer.
(iii) A family of evolutionary sequences starting mass transfer from an
evolved donor with varying $\xc < 0.5$ and low mass transfer rate
($\mdot \simle 5 \times 10^{-9} \msolyr$) covers the observed locations of 
CVs in
the $P-$SpT diagram for $P \simgr 6$ h.   
(iv) For these sequences, higher mass transfer rates than in (i) are
required near 3 h, otherwise they would evolve into the period gap
and predict too early spectral types at shorter $P$.

If $\dot M$ is sufficiently large, the sequences reach their minimum
period above the upper edge of the period gap. The donor
becomes fully convective at a mass smaller than the canonical $0.21
\msun$ of the standard (unevolved) sequence, but at a period longer
than 3~h. If the system detaches at this point and the donor can
re--establish thermal equilibrium, then mass transfer 
would resume at a period well below 2~h. This guarantees
consistency with the standard discontinuous orbital braking model.

Points (ii) to (iv) show that if the mass transfer rate increases
significantly along the secular evolution of evolved donors,
these sequences can explain the observed scatter in the 
$P-$SpT diagram {\it both above} 6~h, 
through the effect a lower $\xc$ has on the evolutionary track (see \S
3 and Fig. \ref{fig3}), {\it and below} 6~h, through the effect of a
higher $\dot M$ (see \S 2 and Fig. \ref{fig1}).  

A large range of the secular mean mass transfer rate, as suggested by (ii),  
could be reconciled with the sharp boundaries of the observed
period gap. 
In the standard model these are a result of the dominance of
sequences with the canonical value $1-2 \times 10^{-9} \msolyr$ at 
periods close to 3~h. 
The apparent scatter of data in the $P$-SpT diagram does not support
this dominance. Reasonably sharp boundaries could be maintained if 
the canonical value is the minimum $\dot M$ value of the range of
transfer rates. The observed gap edges would then be defined by
this minimum mass transfer rate sequence. 
Sequences with higher $\dot M$ would bounce before
they reach the upper edge of the gap. They detach at longer periods
and reattach at shorter periods than the standard sequence.

A more serious problem is that the large $\dot M$ range for  
otherwise {\it similar} system parameters, as suggested by (ii) for
unevolved sequences, leaves the nature of  
the main control parameter determining the strength of orbital 
angular momentum losses completely
undetermined. Our experiments show that $\xc$ could be this parameter. 

The full observed range of data in the $P$-SpT diagram could be
explained if the orbital braking strength increases both with 
decreasing $P$ and decreasing $\xc$. This ensures that period bounce
prevents evolved sequences 
from overpopulating the period gap, while they still pass through
rather late spectral types at long periods. The resulting $\dot J$ law
must give the canonical value of $\dot M$ at 3~h for $\xc = 0.7$. A 
large $\dot M$ range for unevolved systems is not required.


With standard magnetic braking laws (e.g.\ by Verbunt \& Zwaan
1981, Mestel \& Spruit 1987) the transfer rate usually increases with 
period, and is 
smaller for sequences with more evolved donors (see e.g.\ Pylyser \&
Savonije 1989, Singer et al.\ 1993, Ritter 1994), i.e.\ has just the
opposite differential behaviour than the one we propose.
Given the significant uncertainty in these magnetic braking models
and our poor knowledge of the underlying stellar magnetic dynamo, 
the suggestion that observations favour a non--standard law 
does not appear unrealistic. Further considerations 
of the theory of magnetic braking to assess our finding are clearly
beyond the scope of this paper. We note, however, that there are
magnetic braking scenarios in the literature that lead to an increase
of $\dot M$ with progressing evolution. Kolb \& Ritter (1992) obtained 
this for a Verbunt \& Zwaan (1981)--type law when only 
the donor's convective envelope is coupled to the orbital motion.
Zangrilli et al.\ (1997) found this property for their boundary--layer
dynamo.

At this point we re--emphasize that the input physics on which our stellar 
models are based provides an excellent description of observed properties of
isolated low--mass stars. Unless there are effects peculiar to CV
secondaries that are not taken into account by the models (e.g.,
effects due to the rapid rotation, or the irradiation by the white dwarf 
and accretion disc), one is forced to accept that a large fraction of CVs have a 
nuclearly evolved donor star. 
This is in stark contrast to
predictions of standard models for the formation of CVs, where most
CVs form with a 
secondary which is too low--mass ($\la 0.6 \msun$) to be evolved
(Politano 1996, King et al.\ 1994). But this predominance of unevolved
donors is mainly due to the neglect of systems forming with a 
secondary that is significantly more massive than the white
dwarf (Ritter 2000). These undergo thermal--timescale mass transfer at a rate $\ga
10^{-7} \msolyr$
and are usually associated with supersoft sources
(e.g.\ di~Stefano and Rappaport 1994). 
It is perfectly possible that these systems reappear as 
standard CVs once the mass ratio ($q=$~donor
mass/WD mass) is sufficiently small. In descendants of this
thermal--timescale evolution any degree of nuclear evolution prior to
mass transfer is possible. De~Kool (1992) finds that the impact of
these survivor systems depends on the assumed initial mass ratio 
distribution which ultimately determines the secondary  mass
distribution in post--common envelope binaries.
For de~Kool's model 1 (d$N\propto$d$q$) survivor systems completely
dominate the CV population.
A full appraisal of the viability of the apparent predominance of 
nuclear evolved systems is beyond the scope of our study.
This requires a full population synthesis with self--consistent
treatment of $\xc$--dependent angular momentum losses. 

Finally, the accurate determination of donor masses could provide an  
observational test of our hypothesis that a significant fraction of CV
donors is nuclearly evolved. Observational errors are still too large  
(e.g.\ Smith \& Dhillon 1998) for this test. Systems with the most discrepant
(late) spectral type should have the smallest mass at any given $P$.

\paragraph*{Acknowledgments} 

We are grateful to K.~Beuermann, H.~Ritter, H.~Spruit, F.\ and E.~Meyer
for valuable discussions. 
We thank the referee, R.~Smith, for a careful reading of the
manuscript.
I.B thanks the Max--Planck--Institut f\"ur
Astrophysik (Garching) and
the University of Leicester for
hospitality during the realization of part of this work. 
The calculations were performed on the T3E at Centre
d'Etudes Nucl\'eaires de Grenoble.  We thank A.~Norton for improving
the language of the manuscript.

\end{document}